\documentclass{article}
\usepackage{amsmath}
\usepackage{graphicx}
\def\p{\partial}

\def\g{\gamma}

\def\de{\delta}
\def\D{\Delta}
\def\De{\Delta}
\def\ov{\overline}
\def\ld{\lambda}
\def\Ld{\Lambda}

\def\om{\omega}
\def\Om{\Omega}
\def\rh{\rho}

\def\b{\beta}

\def\a{\alpha}

\def\pdellx'{\frac{\partial}{\partial x'}}
\def\pdellw'{\frac{\partial}{\partial w'}}

\newcommand{\be}{\begin{equation}}
\newcommand{\ee}{\end{equation}}
\def\bed{\begin{displaymath}}
\def\eed{\end{displaymath}}
\def\bea{\begin{eqnarray}}
\def\eea{\end{eqncrray}}
\def\[{$$}
\def\]{$$}
\begin{document}
\title{\Large\bf A Generalization of Gauge Symmetry, Fourth-Order Gauge Field Equations and Accelerated Cosmic-Expansion }
\author{ Jong-Ping Hsu\footnote{e-mail: jhsu@umassd.edu}\\
Department of Physics,
 University of Massachusetts Dartmouth \\
 North Dartmouth, MA 02747-2300, USA}
\maketitle
{\small  A generalization of the usual gauge symmetry leads to fourth-order gauge field equations, which imply a new constant force independent of distances.  The force associated with   the new $U_1$ gauge symmetry is repulsive among baryons.  Such a constant force based on baryon charge conservation gives a field-theoretic  understanding of the accelerated cosmic-expansion in the observable portion of the  universe dominated by baryon galaxies.   In consistent with all conservation laws and known forces, a simple rotating `dumbbell model' of the universe is briefly discussed.}
\bigskip	

\section{Introduction}

We propose a generalization of the usual gauge symmetry such as internal $SU_N$ group.  The new $SU_N$ transformations involve a Lorentz vector gauge function $\om^a_\mu(x)$ rather than the Lorentz scalar function $\om^a(x)$ associated with the $SU_N$ group generators $L^a$.  Generalized gauge invariant Lagrangian involves second-order derivatives of gauge fields and new gauge fields satisfy fourth-order differential equations.  Nevertheless, matter fields such as those of fermions and scalars satisfy the usual second-order equations. The phase factor in the usual gauge transformation of, say, a fermion field is generalized to become a non-integral phase factor involving the vector gauge function, $\int^{x} \om_\mu (x') dx'^\mu$, where $\om_\mu(x)=\om^a_\mu(x)L^a$.    In a special subset of gauge functions, in which the vector gauge functions can be expressed as the space-time derivative of scalar functions $\om^a(x)$, i.e., $\om^a_\mu(x)=\p_\mu \om^a(x)$, the non-integral phase factor is simplified to the usual phase factor and the generalized gauge symmetry reduces to usual gauge symmetry.  In this sense, the generalized gauge symmetry contains the usual gauge symmetry as a special subset of vector gauge functions.
 We noted that Lorentz vector gauge functions already appeared in the gauge transformations of the external space-time translational gauge symmetry in Yang-Mills gravity based on flat pace-time.\cite{1,2}

 In 1955, Lee and Yang suggested that there was a $U_1$ gauge field associated with the baryon number (or charge) conservation law.   Their theory is formally identical to the electromagnetic theory.  They used E$\ddot{o}$tv$\ddot{o}$s experiment to estimate the coupling strength between nucleons and found the new inverse-square force to be much weaker ($\le 10^{-5}$) than the gravitational force.\cite{3}  Thus, there is no observable physical effects whatsoever related to such an extremely weak baryon charge. 

However, a modified Lee-Yang gauge field with the fourth-order field equation was discussed in 2005 within the framework of the usual baryonic $U_{1b}$ gauge symmetry with a new Lagrangian involving the quadratic form of the derivatives of gauge-invariant curvature, $\p^\a b_{\mu\nu}$.\cite{4}  It was motivated by the desire to understand the accelerated expansion of the universe based on the baryonic gauge field associated with the established baryon conservation law rather than based on the usual non-field-theoretic and ad hoc `dark energy.'\footnote{As noted by J. M. Keynes: `The difficulty lies not so much in developing new ideas as in escaping from old ones.'}  However, the modified Lagrangian of the baryonic gauge field with the usual baryonic $U_{1b}$ symmetry is not unique, as we shall see below in section 7. 

Therefore, it is natural to ask whether  there is a new gauge symmetry which can `uniquely' specify an invariant Lagrangian in analogy to the usual gauge symmetry, so that we also have a gauge symmetry foundation for the fourth-order gauge field equations.  The answer turns out to be affirmative and this new gauge-symmetric foundation appears to be logically as good as the foundation for  the usual gauge fields with the second-order wave equations.  

In the literature, there were many discussions related to higher-order field equations and their quantizations.  For example, field equations derived from Lagrangians with higher derivatives were investigated by J. S. de Wet in 1948.\cite{5}  He showed how such field equations can be put into Hamiltonian form and how the quantization for boson and fermion fields can be carried out.  He established that the quantization is relativistically invariant and consistent with the field equations.  However, he found that the Hamiltonian proves to be different, in general, from the integral of the 0-0 component of the energy momentum tensor.
Pais and Uhlenbeck also discussed Lagrangians involving higher order derivatives.\cite{6}  They found that there is no essential problem, except that there is non-definite energy for the dynamics system.   Various aspects of higher-order field equations were discussed by many physicists in the past.\cite{7,8,9} 
 But, it seems that there is no connection to basic physical fields\footnote{In mechanical vibration, the beam displacement could be described by fourth order differential equations.  The author wishes to thank E. C. G. Sudarshan for a comment.}  and no experimental predictions based on higher-order field equations, except that some physicists expressed  the hope that higher-order field equations may help to eliminate ultraviolet divergences in quantum field theory.  Such a hope turned out to be difficult to realize because once the ultraviolet divergences are eliminated in this way, the unitarity of the S-matrix will be upset in general so that the probability for a certain physical process may be negative and the theory becomes unphysical.  
 
 For decades, it has been thought that higher-order field equations involve ghost states with non-definite energy or non-positive norm, which will upset unitarity of the theory.   However, it was shown recently that this is not necessarily so,\cite{10,11,12} and thereby one could apply fourth-order field equations to discuss physical phenomena.  Bender and Mannheim proved a no-ghost theorem\cite{10} for the fourth-order derivative Pais-Uhlenbeck oscillator model.  The model represents two oscillators coupled by a fourth-order equation of motion, $d^4 z/dt^4 + (\om^2_1 + \om^2_2)d^2 z/dt^2 + \om^2_1 \om^2_2 z=0$.  With the identification $\om_1 = \om_2 = |{\bf k}|$, this equation is the quantum-mechanical limit of the free field equation $(\p^\a\p_\a)^2 (\p^\b \p_\b)^2 \phi = 0$ in the field configurations of the form   $\phi=z(t) exp(i{\bf k}\cdot {\bf r})$.  This free field equation is formally related to the gauge field equation (14) with $g'=0$ below.  The Bender-Mannheim no-ghost theorem makes field theories with fourth-order field equations interesting and worthwhile further investigation.  In this connection, the fourth-order gauge field equation discussed in this paper is particularly interesting because the new gauge field couples to fermions with baryon charge and the Dirac equations for fermions assure positive energy, ground state and stability of such a fermion system.

The new generalized gauge symmetry appears to be a deeper lying symmetry principle because it contains the usual gauge symmetry, Abelian or non-Abelian, as a special subset of gauge functions (i.e., $\om^a_\mu(x)=\p_\mu \om^a(x)$ ) and it predicts a new kind of cosmic force  independent of distance.  Let us use `taiji' gauge fields\footnote{In ancient Chinese thought, the word `taiji' denotes the ultimate principle or the condition that existed before the creation of the world.} to denote the new gauge fields with the fourth-order differential equations and to distinguish them from the usual gauge fields (with the second-order equations) and from other fields with higher-order equations.

\section{Taiji $U_1$ Gauge Symmetry}

Let us first consider the taiji $U_1$  gauge symmetry of a physical system with gauge field $b_\mu(x)$ and spin 1/2 field $\psi(x)$ (e.g., quarks or leptons).  The taiji gauge transformations for $b_\mu(x)$ and $\psi(x)$ are assumed to be
\be
b'_\mu(x) = b_\mu(x) + \Ld_\mu (x), \ \ \ \ \ \ \  \mu=0,1,2,3, 
\ee
\be
\psi'(x) =   \Om(x) \psi(x), \ \ \ \ \ \ \  \ov{\psi'}(x) = \ov{ \psi}(x)  \Om(x)^{-1},
\ee
\be
   \Om(x)= exp\left(- ig_b \int^{x} \Ld_\ld(x') dx'^\ld \right),
\ee
where $\Om(x)$ is a non-integral (or path-dependent) phase factor, and $g_b$ denote a super weak coupling of the baryon (or the lepton) coupling strength.  The path in (3) could be arbitrary, as long as it ends at the point $x\equiv x^\nu$, and  $\Ld_\mu (x)$ may be considered as space-time dependent parameter of the Lie group $U_1$  and is restricted by the condition
\be
\p^2 \Ld_\mu(x) - \p_\mu \p^\ld \Ld_\ld(x) = 0, 
\ee
$$
\ \ \ \ \ \  \p^2=\p_\mu \p_\nu \eta^{\mu\nu},   \ \ \ \ \   \eta^{\mu\nu}=(1,-1,-1,-1).
$$

As usual, the taiji $U_1$ gauge covariant derivative  $\De_\mu$ is defined as
\be
\De_\mu= \p_\mu + ig_b b_\mu.
\ee
The $U_1$ gauge curvature $b_{\mu\nu}$ is given by
\be
[\De_\mu, \D_\nu]=ig_b b_{\mu\nu},
\ee
\be
b_{\mu\nu} = \p_\mu b_\nu - \p_\nu b_\mu.
\ee
It follows from equations (1)-(4), we have the following taiji gauge transformations for $b_{\mu\nu}(x)$, $\p^\mu b_{\mu\nu}(x)$ and $\De_\mu \psi (x)$:
\be
b'_{\mu\nu}(x)= b_{\mu\nu}(x) +\p_\mu \Ld_\nu (x) - \p_\nu \Ld_\mu (x) \ne b_{\mu\nu}
\ee
\be
\p^\mu b'_{\mu\nu}(x)= \p^\mu b_{\mu\nu}(x) +\p^\mu\p_\mu \Ld_\nu (x) - \p^\mu\p_\nu \Ld_\mu (x)  =\p^\mu b_{\mu\nu}(x),
\ee
\be
  \De'_\mu \psi' (x) =   \Om(x)  (\p_\mu +ig_b b_\mu (x) )\psi(x)= \Om(x)  \De_\mu \psi (x) ,
\ee
where we have used the restriction (4) and
$$
\p_\mu  \Om(x) =  - ig_b   \Ld_\mu (x) \Om(x)
$$
to obtain the results (9) and (10).  The result (8) shows that the $U_1$ gauge curvature $b_{\mu\nu}$ is no longer invariant under the taiji gauge transformations (1), in contrast to the usual $U_1$ gauge symmetry.  However, the space-time derivative of the gauge curvature, i.e., $\p^\mu b_{\mu\nu}$, is invariant under the new gauge transformations (1).  

Furthermore, based on equations (1), (2) and (10), one can show that the usual coupling terms between baryon gauge field $b_\mu(x)$ and the fermion field $\psi$, i.e., $\ov{\psi}\g^\mu\De_\mu \psi (x)=\ov{\psi}\g^\mu(\p_\mu +ig_b b_\mu(x)) \psi (x) $, is taiji gauge invariant: 
$$
\ov{\psi'}(x)\g^\mu \De'_\mu \psi' (x) =\ov{\psi}(x)\g^\mu\De_\mu \psi (x).
$$
Thus, the taiji gauge invariant Lagrangian $L_{b\psi}$ for a system involving fermions (or baryons) $\psi$ and bosons $b_\mu$ is 
\be
L_{b\psi} = \frac{L_b^2}{2}(  \p^\mu b_{\mu\ld} \p_\nu b^{\nu\ld}) +i \ov{\psi}(x)\g^\mu\De_\mu \psi (x)- m\ov{\psi}\psi,
\ee
where $L_b$ is a constant scale with the dimension of length, so that $b_\mu$ has  the same dimension as  the usual vector fields, which satisfy the usual second-order  field equations.  One may consider (11) as the gauge invariant Lagrangian for the 'baryodynamics.'  We note that the modified higher-order Lagrangian, $\propto \p_a b_{\mu\nu} \p^\a b^{\mu\nu}$, for baryonic gauge fields in a previous paper\cite{4} is not invariant under the taiji $U_1$ gauge transformations (1), in contrast to that in the Lagrangian (11) of baryodynamics.

In baryodynamics, the field equations of the taiji gauge field $b_\mu$ and the fermion $\psi$, which carries baryon number (or baryon charge), can be derived from (11).  We have
\be
\p^2 \p^\ld b_{\ld\mu} - \frac{g_b}{L_b^2}\ov{\psi} \g_\mu \psi = 0,
\ee
and 
\be
(i\g^\mu [ \p_\mu + ig_b b_\mu] - m) \psi = 0, \ \ \ \ \ \ \    i [ \p_\mu - ig_b b_\mu] \ov{\psi}\g^\mu+ m\ov{ \psi }= 0.
\ee
The fourth-order equation (12) is invariant under the taiji gauge transformations (1) and may be called taiji gauge field equation.

\section{Linear potentials between baryon galaxies}

Suppose we chose a gauge condition $\p^\ld b_\ld =0$, (12) leads to the following field equations
\be
\p^2 \p^2 b_\mu   = g'_b \ov{\psi}\g_\mu \psi, \ \ \ \ \ \ \ \ \ \   g'_b = \frac{g_b}{L^2_s}.
\ee
Suppose one puts a point-like baryon charge at the  origin.   The zeroth component static potential $b_0({\bf r})$ satisfies the fourth-order equation,
\be
\nabla^2 \nabla^2 b_0 ({\bf r})= g'_b \de^3({\bf r}).
\ee
To solve this equation, we define
\be
b_0({\bf r})  = \int \frac{1}{(2\pi)^{3/2}} d^3 k e^{i{\bf k}\cdot {\bf r}} \ov{b}_0({\bf k}),
\ee
and substitute (16) in (15).  Suppose we multiply both sides of (15) by $exp(-i{\bf k }\cdot {\bf  r})/(2\pi)^{3/2}$ and integrate over $d^3 x$, we obtain
\be
\ov{b}_0({\bf k})= \frac{g'_b}{(2\pi)^{3/2}}\frac{1}{k^2 k^2}, \ \ \ \ \ \   k=|{\bf k}|.
\ee
From (16 ) and (17), we obtain   a linear potential $b_0({\bf r})$,
\be
b_0({\bf r})  =\int \frac{g'_b}{(2\pi)^{3}} d^3 k e^{i{\bf k}\cdot {\bf r}}\frac{1}{k^2 k^2}=-\frac{g'_b r}{8\pi},
\ee
where we have used the Fourier transformation of the generalized functions. \cite{13}

The coupling constant $g_b$ between the gauge field $b_\mu$ and a baryonic field $\psi$ in (5) is dimensionless, provided $b_\mu(x)$ has  the dimension of length. However, the linear  potential (18) and the associated constant force between two baryonic fermions are characterized by the constant $g'_b=g_b/(L_b^2)$, which has the dimension of $(1/length)^2.$  It is unlikely that future experiments can tell us separate values of $g_b$ and $L_b$ in (18).  The reason is that one can absorb the constant $L_b$  in the Lagrangian (11) into the field $b_\mu$, which satisfies  the fourth-order field equation.  After absorbing $L_b$, the dimensions of $b_\mu$ and $g_b$ in the Lagrangian (11) will be changed separately. Nevertheless the dimension of their product $g_b b_\mu(x)$ stays the same.  Thus, it appears that $L_b$ by itself is not an inherent constant of nature.\cite{14}

\section{Some physical implications of the linear potential}

(A) {\bf Constant force could dominate dynamics in large cosmic scales} 
\bigskip

The linear potential (18) implies a  constant repulsive force between baryonic galaxies, independent of the inter-galaxies distances.  A convenient framework to describe the motion of many galaxies appears to be a four-dimensional symmetry framework that is based on the invariance of physical laws and can accommodate common time for all observers in different reference frames.\cite{15} The equation of motion of a freely moving test particle in the Newtonian limit then becomes\cite{4,16}
\be
\frac{d^2 {\bf r}}{dt^2} = {\bf g} + {\bf g}_{bm},
\ee
where ${\bf g}$ is the gravitational acceleration produced by the distribution of ordinary matter, while the constant  acceleration ${\bf g}_{bm}$ is due to the distribution of baryonic matter.    

It is likely that  the coupling strength of the baryonic gauge field $b_\mu$ is extremely weak, its existence cannot be detected in our solar system or even within our galaxy.  However,  the source of this field is ubiquitous baryonic matter rather   than some mysterious `dark energy'  in the universe.  The force of the potential (18) is a super long-range force and will play a dominate role in a very large cosmic scale. The gauge field theory based on the taiji gauge invariant Lagrangian (11) suggests that the accelerated cosmic expansion could be understood in terms of the cosmic force produced by the taiji gauge field in (12).  

In contrast, if one were to assume a suitable cosmological constant $\ld$ in Einstein's gravitational field equation, one would obtain a solution for a static potential involving the term $+\ld r^2/6$ in the Newtonian limit.\cite{16}  Instead of the constant acceleration in (19), a cosmological constant leads to an $r$-dependent acceleration for the cosmic expansion: ${d^2 {\bf r}}/{dt^2} = {\bf g} + C{\bf r}.$  Thus, this difference between the predictions of the new baryonic gauge theory and the Einstein gravity with a cosmological constant can be tested experimentally.  We note, however, that such a cosmological constant in general relativity is neither in harmony with local field theory nor with the conservation law of energy-momentum.\cite{17,18}  In the Newtonian limit, the cosmological constant $\ld$ appears as a constant source in the universe and produces the quadratic potential field,  $+\ld r^2/6$ for the accelerated cosmic expansion.  In this sense, the cosmological constant appears to be some sort of `new aether' with a constant density everywhere in  the universe,\cite{4,16,18} resembling the old aether of 19th century electromagnetism.
\bigskip

(B) {\bf The virial theorem}
\bigskip

Suppose we consider a single particle moving under a central force with the r-dependent potential,\cite{19}
\be
V=a r^{n+1}.
\ee
The time averages of the kinetic energy $\ov{T}$ and the potential energy $\ov{V}$ are related the following relation
\be
\ov{T}= -\frac{1}{2}\ov{V},  \ \ \   V=ar^{-1}
\ee
for the usual inverse square law forces, i.e., $n=-2$.  However, for the super long-range constant force (i.e., n=0 in (20)), we have the relation
\be
\ov{T}= \frac{1}{2}\ov{V}, \ \ \ \    V=ar.
\ee
This relation will be important for calculating the equation of state for a `cosmic system' of gas baryons.
\bigskip

(C) {\bf The equation of the orbit involving elliptic integral}
\bigskip

With the potential (20), the equation of the orbit can be expressed in the well-known form\cite{19}
\be
\theta = \theta_o - \int^u_{u_o} \frac{du}{\sqrt{2mEl^{-2}- 2mal^{-2}u^{-n-1} - u^2}}, \ \ \ \  u=\frac{1}{r},
\ee
where $E$ and $l$ denote respectively constant total mechanical energy and angular momentum.  For the super long-range constant force, i.e., $n=0$, the equation of the orbit (23) involves elliptic integral.  As a result, the orbit for the linear potential, i.e., (23) with $n=0$, takes the form of an elliptic function rather than the usual elliptic orbits in the solar system.
\bigskip

(D) {\bf A possible model of the universe consistent with all conservation laws}
\bigskip

Let us speculate a simple model of the universe,\cite{20} which is consistent with all the conservation laws of electric charge, baryon number, lepton number, etc.   With limited knowledge about the physical universe, it seems premature to conclude the non-existence of anti-matter.  One possible picture of the universe is a `dumbbell universe' with one end dominated by baryon galaxies (or matter) and the other dominated by anti-baryon galaxies (or anti-matter).\footnote{We ignore electrons, the electron-lepton number and others for simplicity.}   These two gigantic clusters could rotate around a center of mass, attracted by cosmic constant forces of baryonic gauge fields. The constant force is presumably too weak to be observed within our galaxy.  Nevertheless, because the strengths of these forces are independent of distance, they would be larger than the gravitational force at extremely large distances.  One end of the dumbbell is dominated by baryon galaxies.  There is an accelerated cosmic expansion due to the constant repulsive force among baryon galaxies in our observable portion of the universe.  Similarly, the other end of the dumbbell is dominated by anti-baryon galaxies, and there is also an accelerated cosmic expansion due to the repulsive constant force among anti-baryon galaxies. However, the whole universe with these two gigantic clusters will be permanently confined by the attractive constant force (or linear potential) between the baryon cluster and the anti-baryon cluster.  This rotating dumbbell model of the physical universe could be stable due to the presence of the constant attractive force. 
 
\section{Taiji gauge invariance with $SU_N$ group}

Let us extend taiji gauge transformations of the $U_1$ group to the $SU_N$ group.  The taiji gauge fields $H^a_\mu$ associated with $SU_N$ transform according to
 \be
H'_\mu(x) = H_\mu(x) + \om_\mu (x) - if\int^{x}dx'^\ld[ \om_{\ld}(x') , H_\mu(x)],   
\ee
where $\om_\ld(x)$  and $H_\ld(x)$ are defined as follows:
\be
\om_\ld(x)=\om_{a\ld}(x)L_a, \ \ \ \ \ \ \   H_\ld(x)=H_{a\ld}(x)L_a.
\ee
We use  $\om^a_\mu(x)$ to denote an infinitesimal (Lorentz) vector gauge functions.
 For fermions, we define the following infinitesimal gauge transformations,
\be
\psi'(x) = \psi(x)  - i f \int^{x}  \om_\ld (x') dx'^\ld \psi(x),
\ee
\be
\ov{\psi'}(x) =\ov{\psi}(x)+i f \ov{ \psi}(x) \int^{x}  \om_\ld (x') dx'^\ld.   
 \ee

 The $SU_N$ has $N^2 -1$ generators $L_a$, which satisfy the commutation relations\cite{21}
\be
[L_a, L_b]=iC^c_{ab}L_c,
\ee
where $a, b, c=1,2,...., N^2-1.$  As usual, the  $SU_N$ covariant taiji gauge derivative is defined by
\be
\De_\mu=\p_\mu + if H_\mu (x)=\p_\mu + if H_{a\mu} (x)L_a.
\ee
We can verify the following transformations for $\De_\mu \psi(x)$,
\be
\De'_\mu \psi'(x) =\De_\mu \psi(x) -if \int^{x} dx'^\ld \om_\ld (x') \De_\mu \psi(x).
\ee

As usual, the $SU_N$ gauge curvature $H^a_{\mu\nu}$ is given by
\be
[\De_\mu, \D_\nu]=if H_{\mu\nu},
\ee
\be
H_{\mu\nu}(x) = \p_\mu H_\nu(x) - \p_\nu H_\mu(x) + if [H_\mu(x), H_\nu(x)],
\ee
or
\be
H_a^{\mu\nu}(x) = \p^\mu H_a^\nu(x) - \p^\nu H_a^\mu(x) - f C_{acd}H_c^\mu(x)H_d^\nu(x).
\ee
We have seen that the taiji gauge curvature $H^a_{\mu\nu}(x)$ for $SU_N$ group is formally the same as the usual gauge curvature.  However, one can verify that, in general,  $H^a_{\mu\nu}(x)$ does not transform properly under the taiji gauge transformation, as one can see from the following transformations,
\be
H'_{\mu\nu}(x)= H_{\mu\nu}(x) +\p_\mu \om_\nu (x) - \p_\nu \om_\mu (x) -if\left[\int^{x} \om_\ld(x') dx'^{\ld}, H_{\mu\nu}(x)\right],
\ee
where $H_{\mu\nu}=H^a_{\mu\nu}L_a $, and we have used equations (24), (25) and (28)).  In the special case, $\om_\mu=\p_\mu \om$, equation (34) reduces to the usual gauge transformations, 
$$
H'^{\mu\nu}_c= H^{\mu\nu}_c +fC_{cab}\om_a H_b^{\mu\nu}= H^{\mu\nu}_c - if(\om_a L_a)_{cb}H_b^{\mu\nu},  \ \ \ \   C_{cab} = -i(L_a)_{cb}.
$$
Thus, we have seen that $H_c^{\mu\nu}$ transforms according to the adjoint representation, as expected in this case.

However, the space-time derivative of $H_{\mu\nu}$, $\p^\mu H_{\mu\nu}(x)$, has the proper transformation property: 
\be
\p^\mu H'_{\mu\nu}(x)= \p^\mu H_{\mu\nu}(x) -if\left[\int^{x}\om_{\ld}(x') dx'^{\ld}, \p^\mu H_{\mu\nu}(x)\right] ,
\ee
provided the functions $\om_{\ld}(x)$ satisfy the restriction
\be
\p^\mu \{\p_\mu \om_\nu (x) - \p_\nu \om_\mu (x)\} -if[\om^\mu(x) , H_{\mu\nu}(x)] = 0,
\ee
for the general case, $\om_{\mu}(x) \ne \p_{\mu} \om(x)$.  The restriction (36) is similar to the restriction for the space-time-dependent parameters ( i.e., gauge functions) of the Lie group in the usual non-Abelian gauge theory.\cite{22,23}

\section{Taiji gauge invariant Lagrangian and fourth-order gauge field equations}
Thus, the taiji gauge invariant Lagrangian $L_{H\psi}$ for a system of fermion $\psi$ and H-bosons $H_\mu$ is 
\be
L_{H\psi} = \frac{L_s^2}{2}(  \p^\mu H^a_{\mu\a} \p_\nu H^{a\nu\a}) +\ov{\psi}(i\g^\mu\De_\mu-m) \psi,\ee
The field equations of the taiji gauge field $H_\mu$ and the baryon $\psi$ can be derived from (37).  We obtain a simple fourth-order equation for the H-boson,
\be
\p^2 \p^\mu H^a_{\mu\nu} -\frac{f}{L_s^2}\ov{\psi} \g_\nu L^a \psi = 0.
\ee
The fermion equations are given by
\be
i\g^\mu ( \p_\mu + if H_{a\mu}L_a)\psi - m\psi = 0, 
\ee
\be
i\p_\mu\ov{\psi} \g^\mu + \ov{\psi}\g^\mu f H_{\mu a} L_a + m\ov{\psi} = 0.
\ee

Since $H^a_{\mu\nu}$ is anti-symmetric in $\mu$ and $\nu$, the taiji gauge equation (38) implies the continuity equation
\be
\p^\mu(\ov{\psi} \g_\nu L^a \psi )=0,
\ee
associated with the $SU_N$ group.  This equation can also be derived from the fermion equations (39) and (40).

\section{Discussions}

Let us consider gauge symmetry and the non-uniqueness of interactions.  In the usual $U_1$ gauge symmetry, the gauge curvature $b_{\mu\nu}$ is invariant under the $U_1$ gauge transformation, as one can see from (8) with  $\Ld_\mu(x)=\p_\mu \Ld(x)$.  Thus, there are infinitely many $U_1$ gauge invariant Lagrangians, which involve quadratic $b_{\mu\nu}$:
\be
b_{\mu\nu}b^{\mu\nu}, \ \ \ \  \p_\ld b_{\mu\nu} \p^\ld b^{\mu\nu}, \ \ \ \  \   \p^\mu b_{\mu\ld} \p_{\nu} b^{\nu\ld}, \ \ \ \    \p_\rh\p_\ld b_{\mu\nu} \p^\rh \p^\ld b^{\mu\nu}, \ etc.
\ee
Thus, the requirement of the usual $U_1$ gauge symmetry does not uniquely determine the usual gauge invariant Lagrangian, $L_{U1}= - (1/4)b_{\mu\nu}b^{\mu\nu}$.  One has to assume simplicity or the minimal gauge invariant coupling.\cite{24} 

Similarly, there are infinitely many taiji gauge invariant Lagrangians, which involve quadratic $\p^\mu b_{\mu\nu}$: 
\be
\p^\mu b_{\mu\ld} \p_{\nu} b^{\nu\ld},  \ \  \p^\rh \p^\mu b_{\mu\ld} \p_{\rh}\p_{\nu} b^{\nu\ld},  \ \  etc.
\ee
  We also assume simplicity for the quadratic form of the gauge invariant $\p^\mu b_{\mu\nu}$ so that we have the Lagrangian (11) for the baryonic gauge field $b_\mu(x)$.

In analogy to the discussions of  Ogievetski and  Polubarinov,\cite{25} the physical meaning of the taiji $U_1$ gauge invariance under the taiji gauge transformation (1) can be analyzed as follows:   Since the baryon gauge field $b_\mu$ satisfies the fourth order field equation, we first decomposed it  as
\be
b_\mu \equiv b^{(1)}_\mu + b^{(0)}_\mu,
\ee
$$
b^{(1)}_\mu=(b_\mu -  \p_\mu \p^{\nu}\p^{-2} b_{\nu}) , \ \ \ \ \ \ \ \    b^{(0)}_\mu=  \p_\mu \p^{\nu}\p^{-2} b_{\nu}= \p^{-2}\p_\mu \p^{\nu} b_{\nu},
$$
$$
\p^{-2}b_\nu(x) = \int \frac{-1}{k^\ld k_\ld}e^{i k\cdot (x-x')} b_\nu(x') \frac{d^4 k}{(2\pi)^4} d^4 x',
$$
where $b^{(1)}_\mu$ and $b^{(0)}_\mu$ are respectively the physical spin 1 and the unphysical spin 0 part of the baryon gauge field.  One can verify that $b^{(1)}_\mu$ satisfies a constraint, $\p^\mu b^{(1)}_{\mu}=0$, so the four components of $b^{(1)}_{\mu}$ are not all independent and $b^{(1)}_{\mu}$ corresponds to a spin 1 vector field.\cite{26}  Under the taiji gauge transformation (1), we have
\be
(b^{(1)}_\mu )'= (b^{'}_{\mu} - \p^{-2} \p_\mu \p^{\nu} b^{'}_{\nu})=b_{\mu}^{(1)}+\Ld_\mu - \p^{-2} \p_\mu \p^{\nu} \Ld_{\nu}= b^{(1)}_{\mu},
\ee
where we have used the restriction (4).  Therefore, we see that the physical spin 1 components, $b^{(1)}_\mu$, of the massless vector field were unchanged under the taiji $U_1$ gauge transformation (1)  and that only the unphysical spin 0 component, $b^{(0)}_\mu$, was changed.  The observable field strength $b_{\mu\nu}$ in (7) contains only the physical spin 1 component and does not involve the unphysical spin 0 component, 
\be
b_{\mu\nu} =\p_\mu(b^{(1)}_\nu + b^{(0)}_\nu) - \p_\nu (b^{(1)}_\mu + b^{(0)}_\mu) = b^{(1)}_{\mu\nu}, \ \ \ \ \  b^{(0)}_{\mu\nu}=0,
\ee
where we have used (44).  This property also holds in quantum electrodynamics (QED).\cite{26}.  The result (46) helps us to understand why the observable physical results in QED and electromagnetic theory are  gauge invariant.

One advantage of the taiji gauge symmetry is that  the taiji $U_1$ gauge covariant derivative  $\De_\mu$ in (5)  can be easily included in a model\cite{27,28} that unifies all the known forces related to gauge symmetries, including gravity. \cite{1}

We have shown the interesting implication of postulating  the baryon number conservation law to be associated with the generalized gauge symmetry.  The `taiji $U_1$ field' satisfies the fourth-order field equation, which can lead to a weak and constant cosmic repulsive force between observable baryon-galaxies.  Therefore,
the new baryonic gauge field can provide a field-theoretic understanding of the cosmic-scale phenomena such as the accelerated cosmic-expansion in our observable portion of the universe, which is dominated by the baryon-galaxies.  

According to the established fundamental symmetries and conservation laws, the physical universe should have equal amount of matter and anti-matter.  Based on this property and the taiji gauge symmetry of baryon gauge field $b_\mu(x)$,  we discuss a new simple `dumbbell model' of the universe with two rotating gigantic baryon-galaxies and anti-baryon-galaxies clusters. It was estimated that baryons dominated over anti-baryon by billions to one, in our observable portion of the universe.  The other portion is probably too far to be observed with the available apparatus at present.  The usual radiations from anti-baryon-galaxies cannot be distinguished from the ordinary baryon-galaxies.  However, the baryon and anti-baryon annihilation radiation can be distinguished from other radiations.\cite{29} Such annihilation radiation should be produced on the `boundary' between regions occupied by baryon-galaxies and anti-baryon-galaxies.  The rotating dumbbell model\footnote{This model may also be pictured as the `taiji yin-yang diagram', in which the taiji circle is divided into two intertwined dark and light halves to represent yin and yang, which is considered as an old Chinese cosmological imagine.} suggests that such a `boundary' exists and could provide an optical search for the anti-matter in the universe, as discussed by Vlasov.\cite{29}

The author would like to thank L. Hsu for a discussion and a description of a rotating dumbbell model of the universe.  The work was supported in part by the Jing Shin Research Fund of the UMassD Foundation.
\bigskip

\noindent
Note added.  The non-integral (i.e., path-dependent) phase factor for $U_1$ and $SU_N$ transformations in (3) and (24) are to be evaluated by path-ordering each term in the series expansion for the exponential.\cite{18}

\newpage  
\bibliographystyle{unsrt}

\end{document}